\documentclass[fleqn,12pt,twoside]{article}
\usepackage{espcrc1}

\usepackage{graphicx}
\usepackage[figuresright]{rotating}

\newcommand{\AmS}{{\protect\the\textfont2
  A\kern-.1667em\lower.5ex\hbox{M}\kern-.125emS}}
\newcommand{\lb}[1]{#1\!\!\!^-}

\title{Theory of fusion hindrance and synthesis of the superheavy elements}

\author{Yasuhisa  Abe\address[MCSD]{Yukawa Institute for Theoretical
Physics,  Kyoto University,  Kyoto 606-8502  JAPAN}, %
        Bertrand  Bouriquet\addressmark,
        Caiwan  Shen\address{Laboratori Nazionali del Sud, INFN, I-95123 Catania, Italy}
        and
        Grigori  Kosenko\address{Department of Physics, University of Omsk, RU-644077 Omsk, Russia}
}       
\begin{document}

% typeset front matter
\maketitle

\begin{abstract}
The two-step model for fusion reactions of massive systems is briefly
reminded.
By the use of fusion probabilities obtained by the model and of
survival probabilities obtained by the new statistical code, we
predict $xn$ residue cross sections for $^{48}$Ca+actinide systems
leading to superheavy elements with Z=114, 116 and 118.
\end{abstract}

\section{INTRODUCTION}
Since the discovery of the periodicity in chemical elements by
Mendelejeff\cite{bib1}, our knowledge has been 
expanded.  The heaviest element in nature is known to be Uranium with
the atomic number
 Z=92.   Heavier elements than that have been synthesized
artificially.  Thus, it is an intriguing
question how many elements can exist or what is the heaviest element
we can synthesize.   Larger $Z$
values result in the instability due to the larger Coulomb repulsion
which would dominate over
nuclear attraction.  But it is also well-known that closed shells of
nucleonic structure give rise to 
an extra-binding which contributes to the stabilization of atomic
nucleus.  Many attempts have been
made to predict double closed shell nuclei heavier than
$^{208}$Pb\cite{bib2}.   
The magic number for proton next to 82 is predicted to be 114, 120 or
126, depending on the nuclear models employed, while that for 
neutron next to 126 is commonly predicted to be 184.   
These results suggest that there is a stable region in the nuclear
chart far away from that of the known isotopes, which is sometimes
called an island of the superheavy elements (SHE).  
Naturally, enormous experimental efforts have been devoted to answer
to the question last few decades.   Until now, the elements with up to
Z=112 have been synthesized by heavy ion fusion experiments with
$^{208}$Pb target\cite{bib3}, and about even heavier elements with Z=114 and 116
are reported indicative experimental observations with $^{48}$Ca beams
on the actinide targets\cite{bib4}.  
But reaction mechanisms of nuclear fusion of massive systems are not
well known, and thus the experiments have been performed following the
systematics of the experimental data available so far.  
Therefore, it is a very important and urgent subject to develop a
theory which permits us to predict which combination of projectile and
target is favorable for synthesis of an element under consideration,
and at what incident energy residue cross sections are optimized.   
Of course, in order to predict residue cross sections quantitatively,
we have to treat not only fusion processes, but also cooling processes
precisely, because compound nuclei formed by fusion reactions are
excited.   
More explicitly, the residue cross section is given by the following
formula, presuming the statistical theory of the compound nucleus,
                                                                               
\begin{equation}
\sigma_{\mbox{\scriptsize res}}(E_{\mbox{\scriptsize
c.m.}})=\pi{\lb{\lambda}}^2\Sigma_J(2J+1)\cdot 
P^J_{\mbox{\scriptsize fusion}}(E_{\mbox{\scriptsize c.m.}})\cdot P^J_{\mbox{\scriptsize surv}}(E^*),
\end{equation}

\noindent
where $E^*$ is the excitation energy of the compound nucleus and is
equal to $E_{\mbox{\scriptsize c.m.}}+Q$ with $Q$ being the $Q$-value
of the reaction,
and $\lb{\lambda}$ is the wave length divided by $2\cdot\pi$.  
The symbol $J$ is the quantum number of total angular momentum of the
system, as usual. 
$P^J_{\mbox{\scriptsize fusion}}$ and $P^J_{\mbox{\scriptsize
surv}}$ denote the fusion and the survival probabilities, respectively. 
The latter is the probability for surviving against
fission decay and charged particle emissions, which are well described
as statistical decays.  
A new computer program KEWPIE
(Kyoto Evaporation Width calculation Program with tIme
Evolution)\cite{bib5} is constructed so as to minimize the
ambiguities, which is briefly described later.  
The remaining unknown is the nuclear masses of the
superheavy isotopes, or in other words, their shell correction
energies, to which $xn$ residue cross sections are extremely sensitive.

On the other hand, as stated above, there is no commonly accepted
theory for the fusion probability, though it is well-known
experimentally that there is a strong hindrance in fusion reactions of
massive systems\cite{bib6}.   
That is, it does not reach 1/2 at the incident
energy equal to an expected barrier height, and in order for that, a
large amount of energy is required additionally, so-called extra-push
energy\cite{bib7}.  
Two possible explanations have been proposed.  
One is due to the loss of the incident energy by friction which is
supposed to act between the ions of the entrance channel\cite{bib8}.  
This is inferred from the studies on
so-called Deep-Inelastic Collisions (DIC).  
Gross and Kalinowski proposed surface friction model (SFM)\cite{bib9}
which successfully explained the characteristic aspects of DIC.  
The other one is due to dissipation of energy of the collective motion 
which leads the system to the spherical compound nucleus, starting
from the pear-shaped configuration formed by the sticking of the ions
of the entrance channel\cite{bib10}.  
This is conceivable, because there is the strong
dissipation in the fissioning motion, i.e., in an ``inverse'' process.
This is confirmed from the studies of anomalous multiplicities of
neutrons, gamma rays, etc. emitted prior to fission of excited
nuclei\cite{bib11}, with the strong friction given by the one-body
wall-and window formula (OBM)\cite{bib12}.  
Thus, it is orthodox and comprehensive to consider that both
mechanisms are working in the fusion reactions of massive systems.
As schematically shown in Fig.~1, 
two processes undergo in time-sequence, i.e., collision processes are 
up to the contact point of the projectile and the target to stick each
other, and subsequently from the sticked configuration, shape
evolution of the united system starts toward the spherical 
shape over the conditional saddle point.  
Therefore, the fusion process is described by the two steps, and then
the fusion probability $P^J_{\mbox{\scriptsize fusion}}$ is given by a
product of the sticking probability $P^J_{\mbox{\scriptsize stick}}$
and the formation probability $P^J_{\mbox{\scriptsize form}}$;

\begin{figure}[htb]
\vskip-0.5cm
\begin{center}
\includegraphics[height=10.6cm]{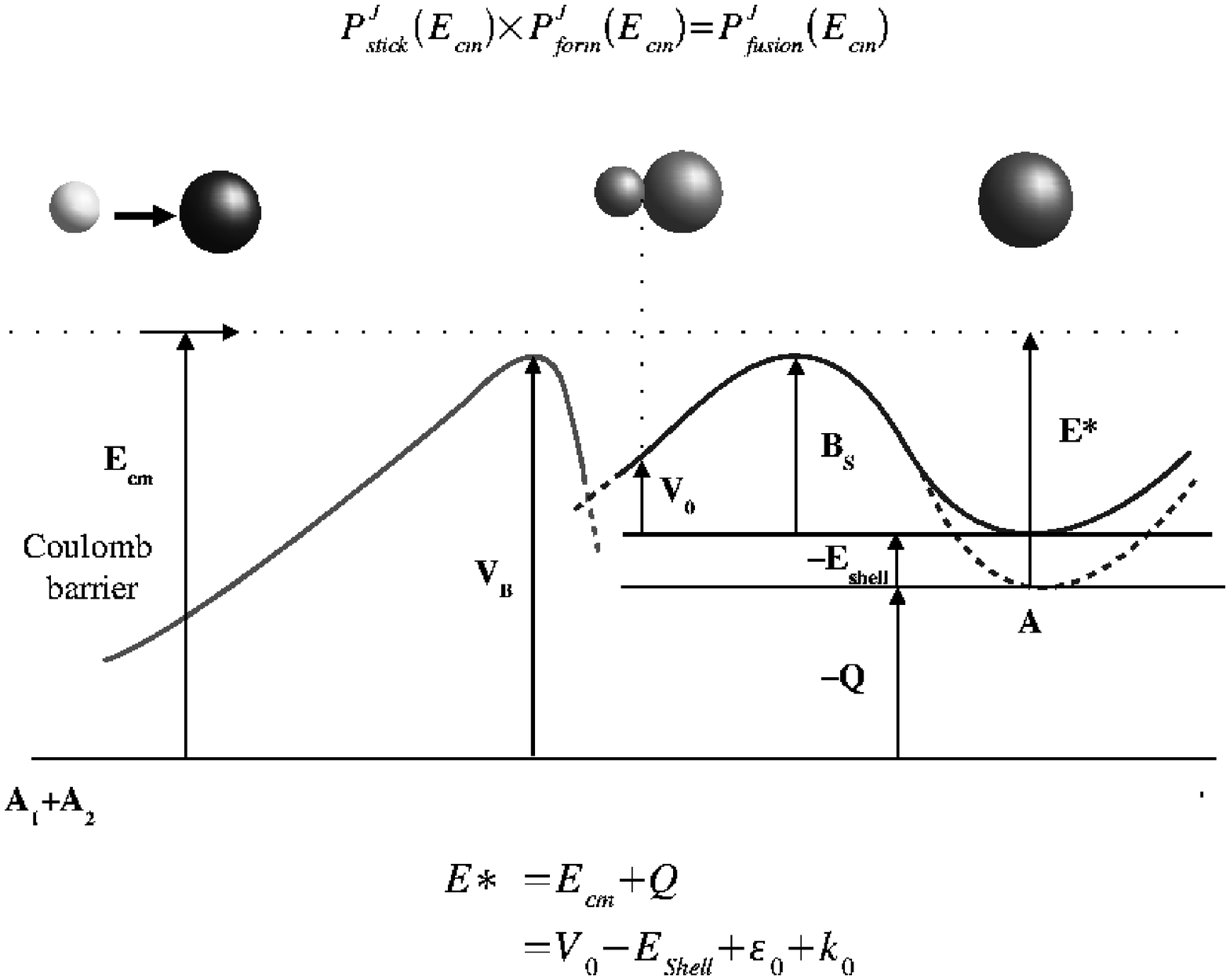}%6.7cm
\vspace{-0.9cm}
\caption{Coulomb barrier, sticking configuration, and conditional
   saddle are shown schematically for massive systems, which
   illustrates a necessity of two-step treatment for fusion.
$E_{\mbox{\scriptsize shell}}$, $\varepsilon_0$ and $k_0$ represent shell 
   correction energy of the spherical shape, intrinsic excitation
   energy and remaining 
   kinetic energy at the contact point.}
\end{center} 
\vskip-1cm
\end{figure}

\begin{equation}
P^J_{\mbox{\scriptsize fusion}}(E_{\mbox{\scriptsize c.m.}})=P^J_{\mbox{\scriptsize stick}}(E_{\mbox{\scriptsize c.m.}})\cdot
P^J_{\mbox{\scriptsize form}}(E_{\mbox{\scriptsize c.m.}})
\end{equation}

This is a new two step model\cite{bib13}.
The method of connection is not like that from a diabatic to an
adiabatic treatments, but should be called ``statistical'', because 
the first step generally results in statistical distributions of
physical quantities of the united system and they should be used as
initial conditions for the second step, as will be detailed below.    

In order to realize the treatment of the first step, SFM is employed,
but in an extended form so as to include the fluctuation forces
associated with the friction forces in accord with the 
dissipation-fluctuation theorem.   
As stated above, the model is used up to the contact point, not for
the whole processes like in DIC.   
Then, the results of the calculations provide the sticking probability
$P^J_{\mbox{\scriptsize stick}}$ for the projectile and the target to
stick each other, as well as information on kinetic energy of the
radial motion, i.e., information on the distribution of the
radial momentum at the contact point.   Since they were already
reported in detail elsewhere\cite{bib14}, we do not repeat it here,
but summarize the results.   

\begin{enumerate}
\item An energy dependence of the sticking probability shows that a
barrier height is effectively shifted to an 
energy about 10 MeV higher than the original one, which would already
provide an explanation of the extra-push energy.  

\vspace{-0.3cm}
\item A distribution of the radial momentum at the contact point 
has almost exactly a Gaussian form whose center,
i.e., the mean value is equal to zero.  
Its variance is consistent with the temperature calculated from the
energy conservation in average.  

\vspace{-0.3cm}
\item The orbital angular momentum is found to reach the dissipation
limit.
\end{enumerate}

These results indicate the fact that the projectile and the target
stick each other at the contact point to form a united system.   
Therefore, with this distribution as the initial values, we solve
shape evolution of the united mono-nuclear system to obtain the
formation probability, i.e., the probability that the system
overcomes the saddle point to reach the spherical shape, by the use of
the same type of Langevin equation which was used for description of
fission of excited nuclei\cite{bib15}. 

Recently it has been shown analytically that the dissipation manifests
its effects in an extremely enhanced way in over-saddle-point problem
with a schematic one-dimensional model.  This is a modeling of the
fusion reactions of massive systems where the sticked configurations
are located outside of the conditional saddle point and thus the
system has to overcome it in order to reach the spherical shape.
Furthermore, we can obtain a simple expression for the extra-push
energy.   
Since they have been reported elsewhere\cite{bib16}, in the present
paper we concentrate ourselves on realistic calculations of the
formation and then the fusion probabilities by the use of the
liquid drop model (LDM).
In the next section, we describe dynamical evolution of nuclear
shapes over the ridge line with Langevin equations for
the collective variables.    
In section 3, we discuss about $^{48}$Ca- induced fusion
reactions on the actinide targets, on which experimental
fusion excitations are available for comparisons.   
By the use of KEWPIE, we calculate the survival probability, which is 
combined with the fusion probability to give $xn$ residue cross sections
for SHE with Z=114, 116 and 118.   
The results are compared with the experimental data available for the
first two elements.¡¡

\section{FORMATION PROBABILITY OF THE SPHERICAL COMPOUND NUCLEUS}

Shapes of the united nucleus are described by the two-center
parameterization (TCP)\cite{bib17} with the distance between two
mass-centers, the mass-asymmetry, and the neck parameter, while the
deformations of fragments are neglected.  
In view of the extremely strong friction for the neck degree of
freedom on the basis of OBM, the neck motion is expected to be very
slow, which permits us to freeze it in a good approximation.  
Thus, the parameter $\epsilon$ is taken to be 0.8.   
It should be worth noticing that we are here interested only in
fusion, but if we are interested in properties of fission-like
fragments, we have to describe dynamics of the neck degree of freedom
as well as those of deformations of the fragments along decaying
paths.   
The shell correction energy is also neglected in formation process, 
since the united system is already well excited in the preceding
processes of overcoming the Coulomb barrier.  
Therefore, we employ LDM for the potential as well as for the 
inertia-mass tensor for the collective motion.  
As for friction, we employ OBM for the
calculation of the friction tensor.  
Now, we need only initial values to start calculations of trajectories 
with the Langevin equation.  
As stated in the introduction, our calculations of shape evolution
start at the contact point.  
That is, the initial value of the distance is equal to the distance
between the mass centers of the configuration, and that of the 
mass-asymmetry is equal to the mass-asymmetry of the entrance channel.  
As for the conjugate momenta, that for the mass-asymmetry is taken to
be zero, while that for the radial
motion can have various values, due to the Gaussian distribution given
by the collision processes over the Coulomb barrier.   
In order to obtain the formation probability, we have to calculate
probabilities $F^J(p_0, T)$ for the system to overcome the ridge line
among numerous trajectories, starting with various initial momenta
$p_0$, and then, to make an average of $F^J(p_0, T)$ with the weight
of the Gaussian 
distribution of the initial momenta $g^J(p_0, T_c)$ obtained by SFM.    
Therefore, the formation probability is given as follows,   

\begin{figure}[htb]
\begin{minipage}[t]{75mm}
\begin{center}
\includegraphics[width=17pc]{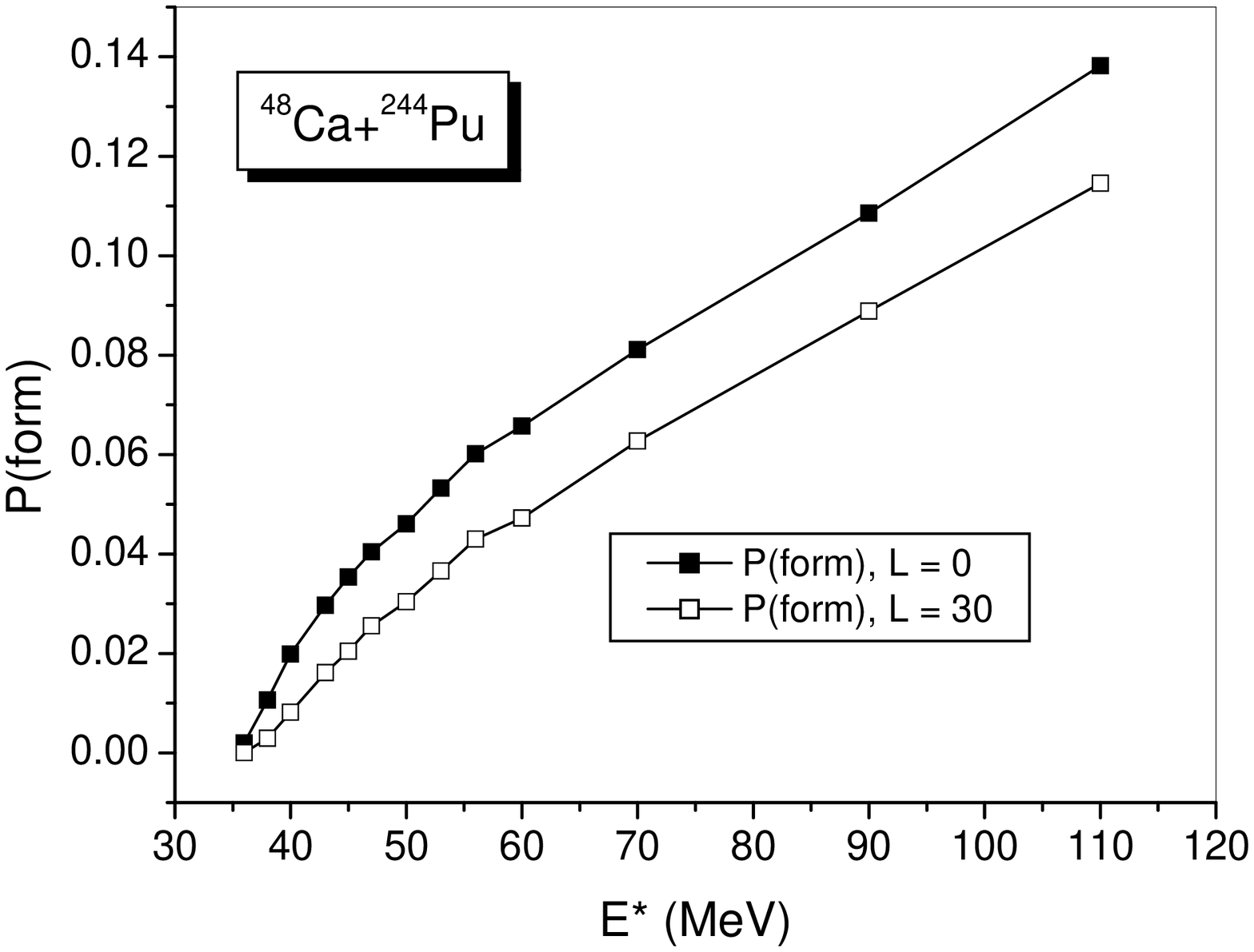}%14pc
\vspace{-0.5cm}
\caption{Calculated formation probabilities for the total angular
momentum $J$ (=incident orbital angular momentum $L$) of 0 and 30.
Note the scale of the ordinate.}
\end{center} 
\end{minipage}
\hspace{\fill}
\begin{minipage}[t]{75mm}
\vspace{-5.5cm}
%\begin{center}
\includegraphics[width=10.7pc,angle=-90]{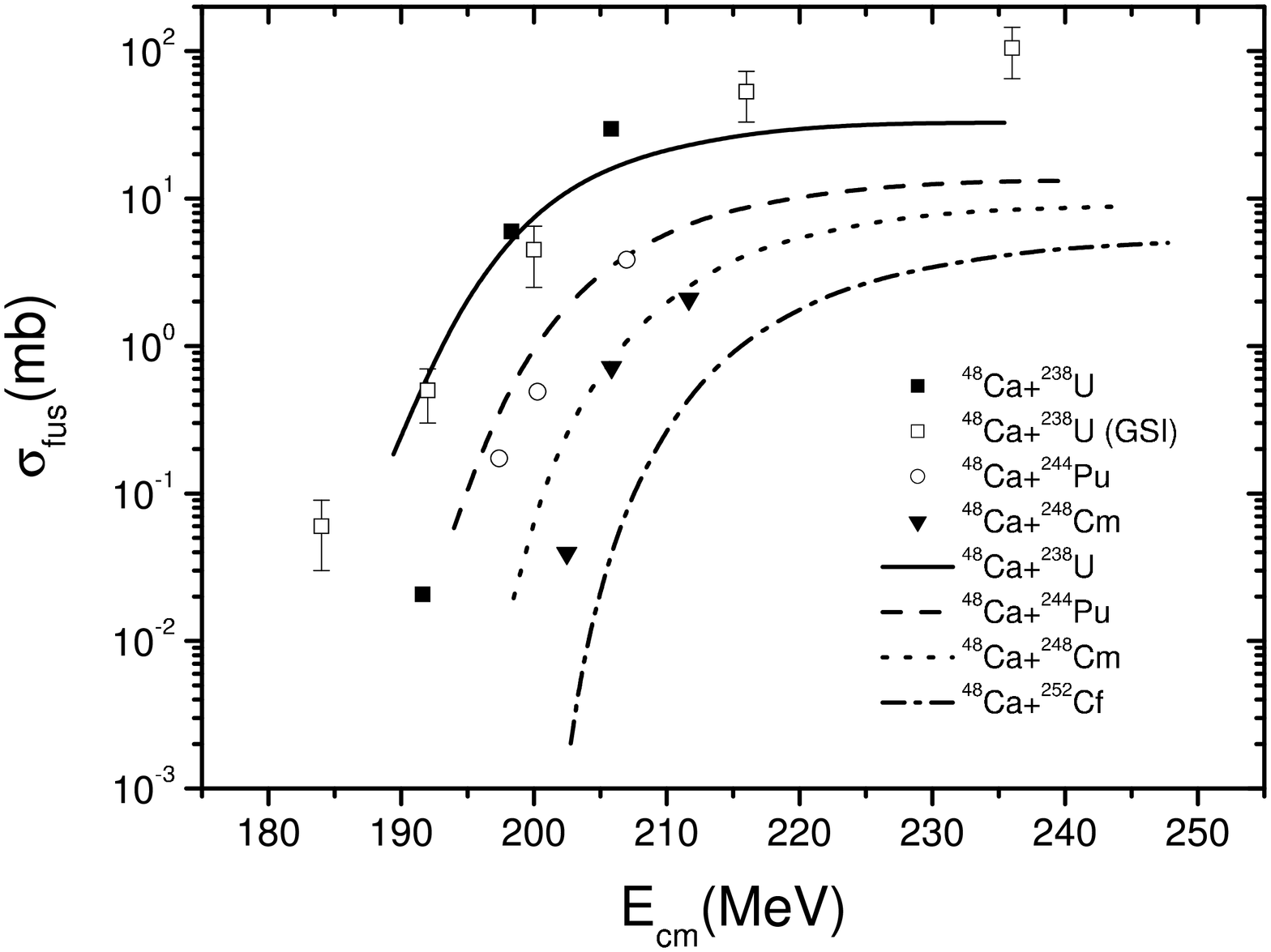}%8.7pc
\vspace{0.3cm}
\caption{Calculated fusion excitation functions for four systems of
$^{48}$Ca + actinide targets are shown, together with the available
experimental data; GSI\cite[a]{bib18} and Dubna\cite[b]{bib18}.}
%\end{center} 
\end{minipage}
\vskip-0.5cm
\end{figure}

\begin{equation}
P^J_{\mbox{\scriptsize form}}(E_{\mbox{\scriptsize c.m.}})=
\int dp_0F^J(p_0,T)\cdot g^J(p_0, T_c),
\end{equation}

\noindent
where $T_c$ denotes the temperature of the system at the contact
point. 
Actually, values of $T_c$ and $T$ are very close each other.
In Fig.~2, the formation probability calculated for 
$^{48}$Ca+$^{244}$Pu system is shown as a function of excitation
energy $E^*$.
It is remarkable that it does not increase quickly as transmission
coefficient usually used for fusion probability, but remains to be
very small over the wide range of energy considered.   
This is in accord with the typical feature of the fusion hindrance
observed.¡¡¡¡

\section{FUSION CROSS SECTIONS AND RESIDUE CROSS SECTIONS OF SHE}

Fusion probabilities are obtained with Eq.~(2).  
Then, fusion excitation functions are calculated as usual,

\begin{equation}
\sigma_f(E_{\mbox{\scriptsize c.m.}})=
\pi{\lb{\lambda}}^2\Sigma_J(2J+1)\cdot P^J_{\mbox{\scriptsize
fusion}}(E_{\mbox{\scriptsize c.m.}})
\end{equation}

The results are shown in Fig.~3 for $^{48}$Ca+ $^{238}$U, +$^{244}$Pu,
+$^{248}$Cm and $^{252}$Cf systems, together with the available
data\cite{bib18}. 
Apparently, it is seen that the calculations reproduce the
experimental feature of the energy-dependence, and furthermore,
reproduce the absolute values of the first 
three systems very well systematically.   
It is worth emphasizing here that there is no adjustable parameter
employed all through the calculations, except the probable choice 
of $\epsilon$.  
Experiments on the last system are strongly desired for confirmation
of validity of the present model.  

From the comparisons made above, it is reasonable to consider that the
fusion probabilities calculated by the present model are accurate, and 
therefore, they can be used for residue calculations with Eq.~(1)
in a high reliability.  
The remaining is to calculate accurately the other factor, i.e., the
survival probability $P^J_{\mbox{\scriptsize surv}}$, which is
made by KEWPIE.
There, it is aimed to reduce ambiguities often introduced
arbitrarily in analyses. 

\begin{enumerate}
\item The level density parameter a is calculated with T\"oke and
Swiatecki's formula\cite{bib19} for the spherical and the saddle-point
shapes.   
For particle evaporations, it is taken to accommodate the shell
correction energy with Ignatyuk's prescription\cite{bib20} with the
standard damping energy of 18.5 MeV.  

\vspace{-0.3cm}
\item As for the saddle point and fission barrier height, we refer to
Cohen-Plasil-Swiatecki's LDM\cite{bib21}.  
But we also adopt the barrier height systematics by Dahlinger et
al. for SHE, which is supposed to be more realistic. 

\vspace{-0.3cm}
\item Kramers factor\cite{bib22} for dynamical fission width is
included with the reduced friction being $5\cdot10^{-20}$/sec which
is consistent with OBM, together with
Strutinski's correction factor\cite{bib23} from inclusion of the collective
degree of freedom around the spherical shape.
\end{enumerate}

The frequencies of the potential at the spherical shape and the saddle
point are temporally taken to be the same, the corresponding energy
quantum taken equal to 1 MeV.  Of course, one could use more realistic
values, but they may give rise to a factor 2 or less.  
Thus, there is no free parameter again in the statistical
calculations, except unknown masses or shell correction energies of
isotopes of SHE.    
As stated in the introduction, there are many structure calculations,
most of which predict shell correction energies of a few to several
MeV.   
M\o ller et al.'s\cite{bib24} are larger ones among them, so we use
their values with reduction factor 1/2 in keeping their tendency in
mass-dependence.  
Calculated excitation functions of $xn$ reactions are shown in Fig.~4
for Z=114, 116, and 118, together with available experimental points.
For Z=114 and 116, the calculations appear not to be inconsistent with
the data\cite{bib4}.   
But the tendency of the predicted masses is not satisfactory.
Nevertheless, it seems that in $^{48}$Ca+$^{248}$Cm system a lower
energy is more favourable.
For the last system, the model provides predictions, on which
experiments are being waited for.

\vskip-1cm
\begin{figure}[htb]
\begin{center}
\includegraphics[width=34.2pc]{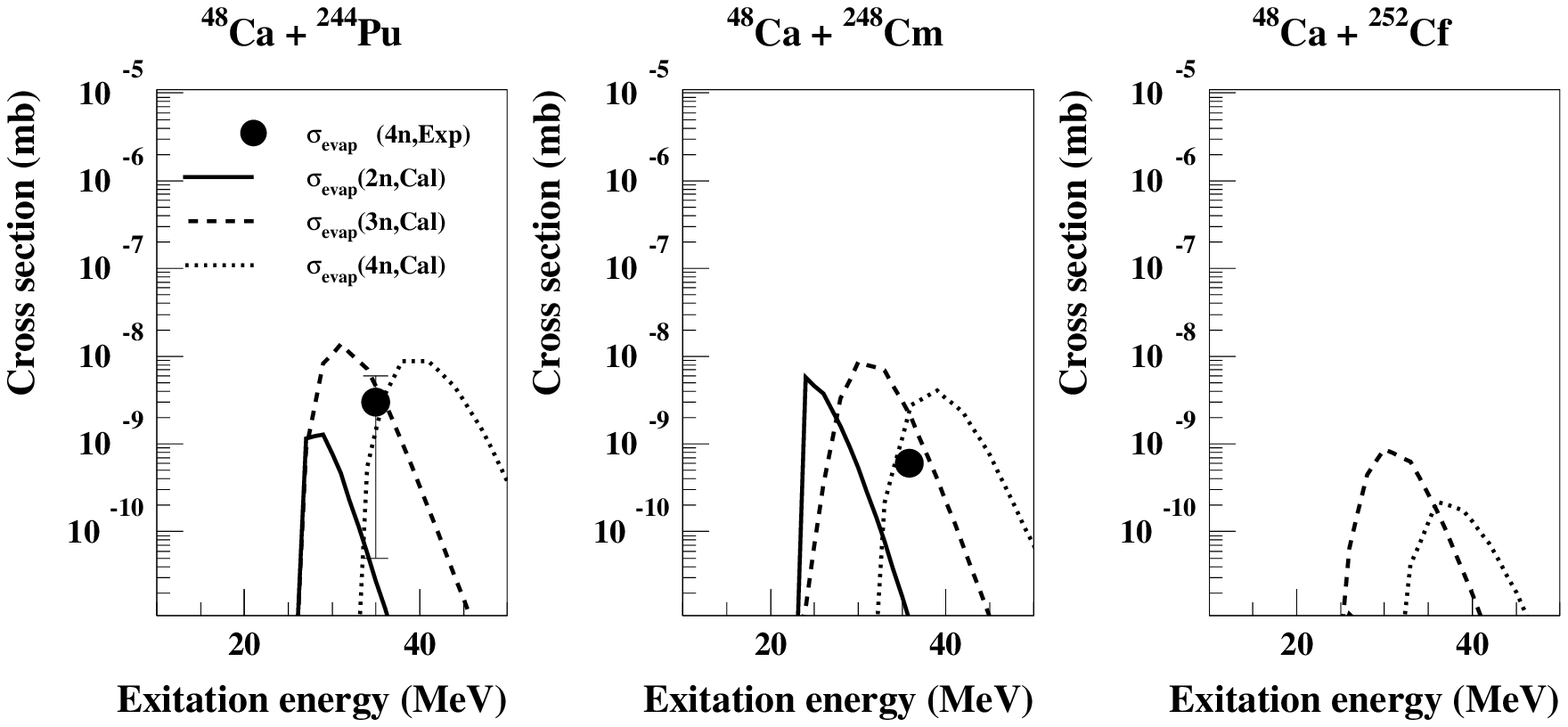}%25pc
\vspace{-1.3cm}
\caption{$xn$ residue cross sections calculated in the present model
are shown for $^{48}$Ca+$^{244}$Pu, +$^{248}$Cm and +$^{252}$Cf systems.
The solid dots are experiments\cite{bib4}.}
\end{center} 
\end{figure}

\vskip-1cm
In brief, the present two-step model has turned out to reproduce the
experimental fusion excitation functions systematically for $^{48}$Ca
induced reactions, as well as the measured $xn$ residue cross sections
for SHE with the shell correction energies of M\o ller et al. with
the reduction factor, the absolute values of which appear to be close
to  some of the mean-field calculations\cite{bib25}.   
Furthermore, we have made predictions on the fusion excitation function
and the $xn$ cross sections for Z=118, which would stimulate
experiments.   
A systematic study is now being made with typical theoretical
predictions of nuclear masses and/or of shell correction energies for
SHE.  
The present model will be applied to other elements and/or other
entrance channels soon.   
Then, we will be able to answer the question raised in the
introduction.

One of the authors (B.B) appreciates a post-doctoral fellowship
provided by JSPS which enables the present collaboration.  
The authors acknowledge communications and encouragements by the
experimental groups at Dubna, GSI, GANIL, RIKEN and JAER\label{}I.
The work is partially supported by a Grant-in-Aids by JSPS
(no. 13640278).


\begin{thebibliography}{9}
\bibitem{bib1}D. Mendelejeff, Z. Chemie 12 (1969) 405.
\bibitem{bib2}a. P.M\o ller and R. Nix, J. Phys. G20 (1994) 1681,\\
b. S. Cwiok et al., Nucl. Phys. A611 (1996) 211,\\
c. M. Bender et al., Eur. Phys. J. A7 (2000) 467.
\bibitem{bib3}S. Hofmann, Rep. Prog. Phys. 61 (1999) 639.
\bibitem{bib4}Yu. Oganessian et al., Phys. Rev. Lett. 83 (1999) 3154
and private communication.
\bibitem{bib5}B. Bouriquet et al., to be published.
\bibitem{bib6}A.B. Quint et al., Z. Phys. A346 (1993) 199.
\bibitem{bib7}S. Bjornholm and W.J. Swiatecki, Nucl. Phys. A391 (1982) 
471.
\bibitem{bib8}P. Fr\"obrich et al., Nucl. Phys. A406 (1983) 557.
\bibitem{bib9}D.H.E. Gross and H. Kalinowski, Phys. Rept. 45 (1978)
175.
\bibitem{bib10}W.J. Swiatecki, Physica Scripta 24 (1981) 113.
\bibitem{bib11}T. Wada et al., Phys. Rev. Lett. 70 (1993) 3538.
\bibitem{bib12}J. Blocki et al., Ann. Phys. (NY) 113 (1978) 330.
\bibitem{bib13}C. Shen et al., Phys. Rev. C66 (2002) 061602(R).
\bibitem{bib14}G. Kosenko et al., J. Nucl. and Radiochem. Sci. 3
(2002) 19.
\bibitem{bib15}Y. Abe et al., J. de Physique 47 (1986) C4-329,\\
Y. Abe et al., Phys. Rept. 275 (1996) Nos.~2 and 3.
\bibitem{bib16}a. Y. Abe et al., Proc. of YKIS01,
Prog. Theor. Phys. Suppl. 146 (2002) 104,\\
\,\,\,\,\, Y. Abe et al., Proc. Rauischholzhausen Meeting, to be published,\\
b. Y. Abe et al., Proc. of Zakopane School 2002, to be published in
Acta Physica\\
 \,\,\,\,\, Polonica B (2003),\\
c. W.J. Swiatecki et al., ibid.
\bibitem{bib17}K. Sato et al., Z. Phys. A290 (1979) 145.
\bibitem{bib18}a. W.Q. Shen et al., Phys. Rev. C36 (1987) 115,\\
b. M.G. Itkis et al., Il Nuovo Cimento 111A (1998) 783.
\bibitem{bib19}J. T\"oke and W.J. Swiatecki, Nucl. Phys. A372 (1981)
141. 
\bibitem{bib20}A.V. Ignatyuk et al., Sov. J. Nucl. Phys. 21 (1975)
255. 
\bibitem{bib21}S. Cohen et al., Ann. Phys. (NY) 82 (1974) 557.
\bibitem{bib22}H.A. Kramers, Physica VII 4 (1940) 284.
\bibitem{bib23}V.M. Strutinski, Phys. Lett. B47 (1973) 121, see also
ref.~15. 
\bibitem{bib24}P. M\o ller et al., Atomic Data and Nuclear Data Tables 
59 (1995) 185.
\bibitem{bib25}T.J. Buervenich, private communication.
\end{thebibliography}
\end{document}